\documentclass{article}

\usepackage{times,amsmath,amsfonts,amssymb,latexsym,textcomp}
\usepackage{color}
\usepackage{graphicx}
\usepackage{multirow}
\usepackage{bbm}
\usepackage{tabularx}
\usepackage{amsthm}
\usepackage{dsfont}
\usepackage{float}
\usepackage{array}
\usepackage{placeins}
\usepackage{MnSymbol}
\usepackage[electronic]{ifsym}
\usepackage[lofdepth,lotdepth]{subfig}
\usepackage{relsize}
\usepackage[T1]{fontenc}
\usepackage[latin9]{inputenc}
\usepackage[english]{babel}

\begin{document}

\title{Interference Energy Spectrum of the Infinite Square Well}
\author{Mordecai Waegell \footnote{caiwaegell@gmail.com},  Yakir Aharonov, Taylor Lee Patti \\ \it{ Institute for Quantum Studies, Chapman University} \\ {\it Orange, CA, USA}}

\maketitle

\textbf{Abstract:} Certain superposition states of the 1-D infinite square well have transient zeros at locations other than the nodes of the eigenstates that comprise them. It is shown that if an infinite potential barrier is suddenly raised at some or all of these zeros, the well can be split into multiple adjacent infinite square wells without affecting the wavefunction. This effects a change of the energy eigenbasis of the state to a basis that does not commute with the original, and a subsequent measurement of the energy now reveals a completely different spectrum, which we call the {interference energy spectrum} of the state. This name is appropriate because the same splitting procedure applied at the stationary nodes of any eigenstate does not change the measurable energy of the state. Of particular interest, this procedure can result in measurable energies that are greater than the energy of the highest mode in the original superposition, raising questions about the conservation of energy akin to those that have been raised in the study of superoscillations. An analytic derivation is given for the interference spectrum of a given wavefunction $\Psi(x,t)$ with $N$ known zeros located at points $s_i = (x_i, t_i)$. Numerical simulations were used to verify that a barrier can be rapidly raised at a zero of the wavefunction without significantly affecting it. The interpretation of this result with respect to the conservation of energy and the energy-time uncertainty relation is discussed, and the idea of alternate energy eigenbases is fleshed out. The~question of whether or not a preferred discrete energy spectrum is an inherent feature of a particle's quantum state is examined.

\section{Introduction}

The purpose of this letter is to show that it is possible, in principle, to measure alternate energy eigenbases of a given superposition state of the infinite square well and that the highest energy eigenstate in a given superposition may have a different energy in different bases.

The origin of this idea goes back to the study of superoscillations initiated by Aharonov \textit{et al}., who first raised the question about extracting a particle from a superposition state with an energy greater than that of its highest mode \cite{aharonov2016super, aharonov2002superoscillations, ferreira2002energy, berry2006evolution, ferreira2006superoscillations, tollaksen2007novel, ferreira2007construction, berry2009natural, berry2012pointer, aharonov2013some, lee2014direct}. Transient zeros of the wavefunction have also been considered in the study of quantum revival and quantum carpets \cite{berry2001quantum,kaplan2000multimode,friesch2000quantum}.

We consider only the 1-D infinite square well, but the findings here can be trivially generalized to the 3-D case. We proceed with the simplest example of the effect in question, and after this, we~give the general derivation for arbitrary superposition states.

The measurement of an alternate energy eigenbasis is performed in two stages. In the first stage, at a moment when there is a zero in the wavefunction, an infinite delta-function potential barrier is suddenly raised at the location of a zero, which has the effect of dividing the original infinite square well into two adjacent infinite square wells, while causing virtually no perturbation to the wavefunction (a similar process is discussed in \cite{potovcek2015quantum}, although with quite a different purpose, and an analysis of perturbation theory with singular potentials is given in \cite{sen1999perturbation}). This division results in a superposition state of the particle being on one side of the barrier or the other and, furthermore, a superposition of the energy levels of each individual well. We call the combined spectra of the two individual wells an {\it interference spectrum}
. This process has effectively accomplished a spectrum and, thus, frequency conversion of the state, which may be quite novel when compared to other related techniques \cite{remez2015super, suchowski2014adiabatic, leshem2014experimental, jain1996efficient, stolen1982parametric, huang1992observation, cerny2004solid, milchberg1995high}.

In the second stage, the energy of the state is measured and is now found in an energy eigenstate of one of the two new wells, rather than an eigenstate of the original well. This is the real effect of raising the barrier: it changes the list of eigenstates onto which the state can collapse when measured.

Of particular interest is the fact that in the new spectral decomposition of the state, it may be possible to measure an energy higher than the energy of the highest mode in the original spectral decomposition of the state. In general, there is no evidence of a violation of conservation of energy because the sudden barrier introduces a large energy uncertainty due to the energy-time uncertainty relation \cite{mandelstam1945uncertainty, aharonov1961time, moshinsky1976diffraction, kobe1994derivation, deffner2011energy}.

A wavefunction that contains a region of superoscillation turns out to be a special case of this phenomenon, wherein very particular superposition states have transient zeros that remain stable for extended \mbox{durations \cite{berry2006evolution}}. Because of the stability of these zeros, barriers can be raised very slowly, and the new spectrum can be obtained without introducing a large energy uncertainty, which may be interpreted as causing a violation of the conservation of energy \cite{aharonov2016super}.

Here, we propose that all we have done through this two-stage process is to effect a measurement in an energy eigenbasis that does not commute with the original energy eigenbasis of the state. The barrier can be raised with virtually no perturbation to the wavefunction, and this has the effect of changing the discrete energy spectrum of that wavefunction. With this interpretation, the wavefunction itself does not have a definite preferred energy spectrum until it is measured with specific boundary conditions (\emph{i.e.}, one spectrum without the barrier or another with the barrier). This~nullifies the issue of a violation of the conservation of energy, since the original spectrum places no special restriction on what energies can be obtained from a measurement.

While the idea of alternative energy eigenbases for the infinite square well may seem somewhat radical, we point out that a spin-$\frac{1}{2}$ particle can be measured in complementary Pauli-spin eigenbases, and each measurement is performed by coupling the spin to a different Hamiltonian. Just as in the present case, the change of spectrum has no effect on the state; all that has changed is the list of eigenstates into which the particle can collapse when a measurement of the energy is performed.

Finally, we have performed extensive numerical simulations of the evolution of several key wavefunctions as a Gaussian barrier of various widths is raised at various rapid speeds to a finite potential. We used the simulation data over this range of parameters to come up with an approximate characterization of how a narrow barrier changes the wavefunction as a function of the barrier's speed, width and the characteristics of the \mbox{initial state}.

Our simulations verify that if a very narrow barrier is raised sufficiently fast at a zero of the wavefunction, the splitting of the well and the change of energy spectrum can indeed be accomplished with virtually no change to the wavefunction.

The remainder of this paper is organized as follows: in the next section, we explore the simplest case of a wavefunction with a single transient zero in complete detail and introduce an example case that might allow experimental verification of these ideas. Next, we discuss the details of raising the barrier; the sudden and adiabatic approximations and the parametric conversion of the spectrum and splitting of the the eigenstates. After this, we discuss a preliminary idea for an experimental implementation of this effect. We then present the formalism for the general interference spectrum of a general superposition state. Finally, we discuss energy conservation and energy-time uncertainty in alternative interpretations of this effect and end the paper with a few concluding remarks. In the Appendix, we discuss the numerical simulations of the time-dependent Schr\"{o}dinger equation that we conducted in order to characterize the effect of rapidly raising a Gaussian barrier.

\section{Results}

\subsection{The Simple Case}

To begin, we will take our infinite square well, which we will call Well 0, to be of width $L$, with boundaries located at $x=0$ and $x=L$. The energy eigenstates of this well are,
\begin{equation}
\psi^0_l(x) = \sqrt{\frac{2}{L}}\sin\frac{l \pi x}{L},
\end{equation}
and have corresponding energies,
\begin{equation}
E_0(l) = \frac{\hbar^2 \pi^2 l^2}{2ML^2}.
\end{equation}

Consider the following normalized superposition of the ground state ($l=1$) and first excited state ($l=2$),
\begin{equation}
\psi(x) = \sqrt{\frac{2}{L (\alpha^2+1)}}\left(\alpha \sin\frac{\pi x}{L} - \sin\frac{2 \pi x}{L}\right),  \label{Psi}
\end{equation}
with $\alpha \equiv 2\cos\left[\frac{\pi x_0}{L}\right]$, and $x_0 \in (0,\frac{L}{2})$ is a zero of $\psi(x)$. This zero is transient and quickly vanishes as the state evolves in time. During a complete period of evolution, this state also develops a transient zero at $x_1 = L-x_0$, and so, we define the list of zero points for $\Psi(x,t)$ as $s = \{(x_0,t_0), (x_1, t_1) \}$.

Thus, at any given time, this function has at most one zero inside the well, and by symmetry, we only need to consider the case of $(x_0, t_0)$. This zero is technically only present at a single instant in time, and thus, the barrier must be raised instantaneously. Clearly, both the delta-function potential and the sudden implementation are the nonphysical ideal case.

Now, suppose that at time $t_0=0$, we raise a new infinite delta-function potential barrier at $x_0$, splitting the original well into two smaller wells of widths $x_0$ and $L-x_0$, which we will call Well 1 and Well 2, respectively. $\psi(x)$ already satisfies the new boundary conditions, and so, there is no instantaneous change in the wavefunction or the expectation value of any observable. The probabilities to find the particle in either well are,
\begin{equation}
P_1=\int_{0}^{x_0} |\psi(x)|^2 dx, \hspace{1cm} \hspace{1cm}  P_2=\int_{x_0}^{L} |\psi(x)|^2 dx.
\end{equation}

Defining the truncated and renormalized wavefunctions in each well as:
\[ \psi_1(x) = \left\{
\begin{array}{ll}
   \psi(x)/\sqrt{P_1} & 0 \leq x \leq x_0 \\
   0 & x_0 < x \leq L \\
\end{array}
\right. \]
and
\[ \psi_2(x) = \left\{
\begin{array}{ll}
   0 & 0 \leq x \leq x_0 \\
   \psi(x)/\sqrt{P_2} & x_0 < x \leq L \\
\end{array},
\right. \]
we can rewrite the original wavefunction as:
\begin{equation}
\psi(x) = \sqrt{P_1}\psi_1(x) + \sqrt{P_2}\psi_2(x).
\end{equation}

If we note that after the barrier goes up, a classical particle must either be in Well 1 or Well 2, we can interpret this wavefunction as a superposition of the particle being in Well 1 in the state $\psi_1(x)$ with probability $P_1$ or in Well 2 in the state $\psi_2(x)$ with probability $P_2$.

Defining,
\begin{equation}
\langle E \rangle = \int_0^L \psi^*(x) \hat{H} \psi(x) dx,
\end{equation}
\begin{equation}
\langle E_1 \rangle = \int_{0}^{x_0} \psi_1^*(x) \hat{H} \psi_1(x) dx,
\end{equation}
and:
\begin{equation}
\langle E_2 \rangle = \int_{x_0}^L \psi_2^*(x) \hat{H} \psi_2(x) dx,
\end{equation}
gives us the relation,
\begin{equation}
\langle E \rangle = P_1\langle E_1 \rangle + P_2\langle E_2 \rangle.
\end{equation}

The state has expectation value $\langle E \rangle$ because with probability $P_1$, the particle is in Well 1 with average energy $\langle E_1 \rangle$, and with probability $P_2$, it is in Well 2 with average energy $\langle E_2 \rangle$.

Wells 1 and 2 have energy eigenstates,
\[ \psi^1_n(x) = \left\{
\begin{array}{ll}
   \sqrt{\frac{2}{x_0}}\sin\frac{n \pi x}{x_0} & 0 \leq x \leq x_0 \\
   0 & x_0 < x \leq L \\
\end{array}
\right\}, \]
and:
\[ \psi^2_m(x) = \left\{
\begin{array}{ll}
   0 & 0 \leq x \leq x_0 \\
   \sqrt{\frac{2}{L-x_0}}\sin\frac{m \pi (x-x_0)}{L-x_0} & x_0 < x \leq L \\
\end{array}
\right\}, \]
respectively, with corresponding energy eigenvalues,
\begin{equation}
E_1(n) = \frac{\hbar^2 \pi^2 n^2}{2Mx_0^2},
\end{equation}
and:
\begin{equation}
E_2(m) = \frac{\hbar^2 \pi^2 m^2}{2M(L-x_0)^2}.
\end{equation}

In general, these energy levels are different from one another and from $E_0(l)$. Furthermore, it is possible to measure an energy $E_1(n)$ or $E_2(m)$ larger than $E_0(2)$, which is the highest mode of Well 0 present in the superposition state $\psi(x)$. This is then an example of a superoscillatory effect.

If we measure the energy of the original well, we will find energy $E_0(1)$ with probability \linebreak $\alpha^2/(\alpha^2+1)$ and energy $E_0(2)$ with probability $1/(\alpha^2+1)$, indicating that we have projected the state onto states $\psi^0_1(x)$ or $\psi^0_2(x)$. If instead, we split the well by putting up the barrier at $x_0$ and then measure the energy, the measurement projects onto states $\psi^1_n(x)$ or $\psi^2_m(x)$.

To find the probability to collapse onto eigenstates of the split well, we decompose $\psi(x)$ into the modes of the split wells:
\begin{equation}
\psi(x) = \sum_{n=1}^\infty a_n \psi^1_n(x) + \sum_{m=1}^\infty b_m \psi^2_m(x)     \label{PsiDecomp}
\end{equation}
with $a_n$ and $b_m$ as shown below:
\begin{equation}
a_n = \frac{2}{\sqrt{x_0 L (\alpha^2+1)}}\mathlarger{\int_{0}^{x_0}}\left(\alpha \sin\frac{\pi x}{L} - \sin\frac{2 \pi x}{L}\right)\sin\frac{n \pi x}{x_0}dx
\end{equation}
\begin{equation}
=\frac{2 n L^{\frac{3}{2}} \sqrt{x_0} (-1)^n }{\pi \sqrt{\alpha^2+1}} \left( \frac{\alpha \sin\frac{ \pi x_0 }{L}} {{x_0}^2 - n^2L^2} - \frac{\sin\frac{2 \pi x_0 }{L}} {4{x_0}^2 - n^2L^2} \right)
\end{equation}
\begin{equation}
b_m =\frac{2}{\sqrt{(L-x_0)L(\alpha^2+1)}}\mathlarger{\int_{x_0}^{L}} \left(\alpha\sin\frac{\pi x}{L}-\sin\frac{2 \pi x}{L}\right)\sin\frac{m \pi (x - x_0)}{L - x_0} dx
\end{equation}
\begin{equation}
= \frac{-2 m L^{\frac{3}{2}} }{\pi}\sqrt{\frac{L - x_0}{\alpha^2+1}}\left(\frac{\alpha\sin\frac{\pi x_0}{L}}{(L-x_0)^2-m^2L^2} - \frac{\sin\frac{2 \pi x_0}{L}}{4(L-x_0)^2-m^2L^2}\right).
\end{equation}

The mod-squared coefficients $|a_n|^2$ and $|b_m|^2$ are then the probabilities to find the particle in each eigenstate. Additionally, of course,
\begin{equation}
\sum_{n=1}^\infty |a_n|^2 + \sum_{m=1}^\infty |b_m|^2 = P_1 + P_2 = 1.
\end{equation}

The two alternate energy eigenbases ($\{\psi^0_l(x)\}$ and $\{\psi^1_n(x),\psi^2_m(x)\}$) each span the space of normalizable functions that are zero at $x_0$, but the corresponding Hamiltonians do not commute; thus, these two energy eigenbases are complementary (or at least, there is some uncertainty relation between them).

We call the new energy spectrum that is available using this measurement procedure the {\it interference energy spectrum of the state} $\psi(x)$. This name is appropriate because the available energies that can be measured are only different from $E_0(l)$ if $\psi(x)$ is a superposition of different $\psi^0_l(x)$.

The simplest way to see this is by considering the state $\psi^0_2(x)$ by itself, which is also obtained by considering the above treatment for $\psi(x)$ in the limit that $x_0 \rightarrow L/2$. This state has a definite energy $\langle E \rangle = E_0(2) = \frac{2 \hbar^2 \pi^2}{ML^2}$ and a stationary zero at $x=L/2$. If we insert an infinite barrier at this zero and split the well, we get equal probability to find the particle in the ground state of each sub-well (\mbox{left or right}), with energy $\langle E_1 \rangle = \langle E_2 \rangle = \frac{2 \hbar^2 \pi^2}{ML^2} = E_0(2)$. Thus, even if we split the well, we always find the particle with the same wavelength and, thus, the same energy. It is trivial to see that this generalizes to splitting any eigenstate at any subset of its nodes.

\begin{figure}[H]
\begin{center}
\includegraphics[width=3.5in]{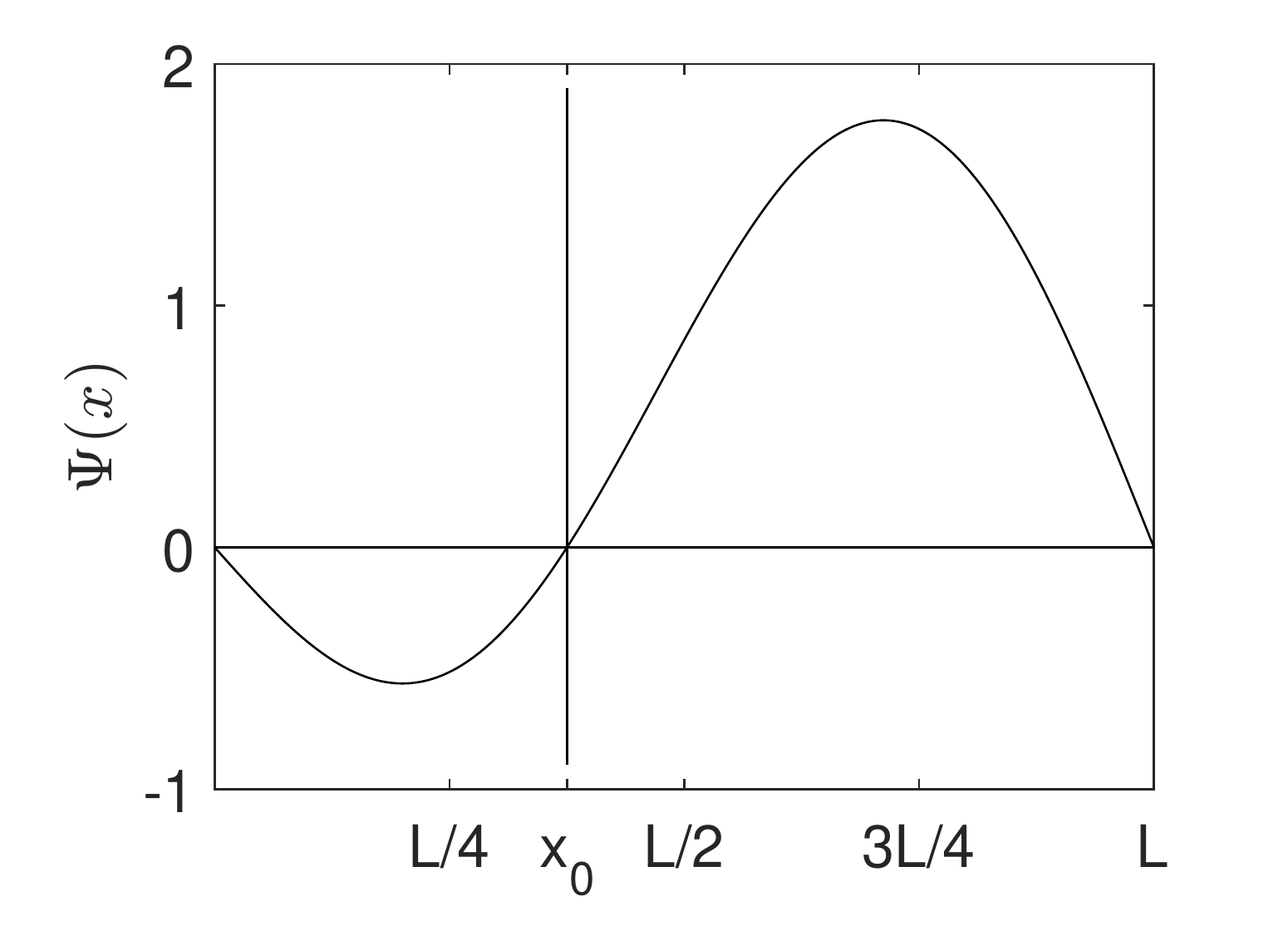}
\end{center}
\caption{Plot of $\psi(x)$ with $x_0 = \frac{3}{8}L$ where a sudden barrier can divide the well.}\label{SpecState}
\end{figure}

Of particular interest for experiments would be to prepare the state $\psi(x)$ with~\mbox{$x_0 = \frac{3}{8}L$}, shown in Figure \ref{SpecState}. If we put up the barrier at $x_0$ and measure the energy, we find that the probability to find the particle in the ground state of the first well, \mbox{$P(n=1) \approx 6\%$}, with energy $E_1(1) = \frac{64}{9}\frac{\hbar^2 \pi^2}{2ML^2}$, which exceeds the energy $E_0(2) = 4\frac{\hbar^2 \pi^2}{2ML^2}$~of~the~highest mode in $\psi(x)$ by a factor of $\frac{16}{9}$. Thus, if the experiment can be performed, it should be possible to measure superoscillatory interference energies with plausible \mbox{success rates}.

For all values of $x_0$, the ground states of either split well are always the most probable, with the probabilities, $|a_n|^2$ and $|b_m|^2$, of measuring higher modes converging toward zero as $n,m \rightarrow \infty$. It is nevertheless possible to measure arbitrarily high energy outcomes with nonzero probability, whereas only the two lowest modes were present in the original well.

\subsection{Raising the Barrier: The Sudden Approximation and the Adiabatic Limit}

We made the assumption above that if the delta-function barrier is raised very quickly at a transient zero of the wavefunction, then the wavefunction itself is not significantly changed by the process, and as a result, the expectation values of all observables are also unchanged.
Because both a delta-function barrier and an instantaneous potential change are nonphysical, we have performed extensive numerical simulations of the time-dependent Schr\"{o}dinger equation with a Gaussian barrier of varying widths $w$, raised linearly to a large finite height over a finite period of time that is very short relative to the characteristic frequencies of the initial states. The simulation was run over a representative range of physically plausible parameters, with emphasis on the narrow-barrier regime.

We used $\psi(x)$ with $x_0 = \frac{3}{8}L$ for the simulation, which was performed using a modified fourth-order Runge--Kutta method. More technical details about the simulation can be found \mbox{in Section \ref{Simulation}}.

The results of the simulation show that as the barrier is made wider, the change to the energy of the state grows smaller and goes to zero, and in the nonphysical limit that the width goes to zero (the delta-function limit), it appears to go to infinity as $~1/w^3$ (see Equations (\ref{DKfit}) and (\ref{DVfit}). This can be overcome in the equally nonphysical limit that the barrier is raised instantaneously, in which case there is no change in energy. We do find that for physically-reasonable barrier widths, final barrier magnitudes and raising periods, the lowest modes of the well can be effectively split with a negligibly small perturbation to the state and its energy. We obviously do not get the exact spectrum we would if the well had been split by a delta-function, but the spectrum and eigenstates are certainly close enough to obtain a superoscillatory energy.

For example, set the width to $w = 10^{-3}[L]
$ and the total period to raise the barrier to a maximum scale of $V=10^4 [\hbar^2/2ML^2]
$ to $\tau = 10^{-10}[2ML^2/\hbar]$. The kinetic energy of the original state is $\langle K \rangle~=~2.8918 [\hbar^2\pi^2/2ML^2]$, and the change in kinetic energy is on the order of $10^{-9} [\hbar^2/2ML^2]$, which is below the error threshold of the simulation. With this barrier, the ground state kinetic energy is $E(1) \approx 2.5605 [\hbar^2\pi^2/2ML^2]$ (this is a numerical result), compared to $E(1) = 2.5600 [\hbar^2\pi^2/2ML^2]$ for the delta-function barrier, and the corresponding wavefunctions are also nearly identical; thus, the desired splitting has been performed.

\begin{figure}[H]
\begin{center}
\includegraphics[width=5.5in]{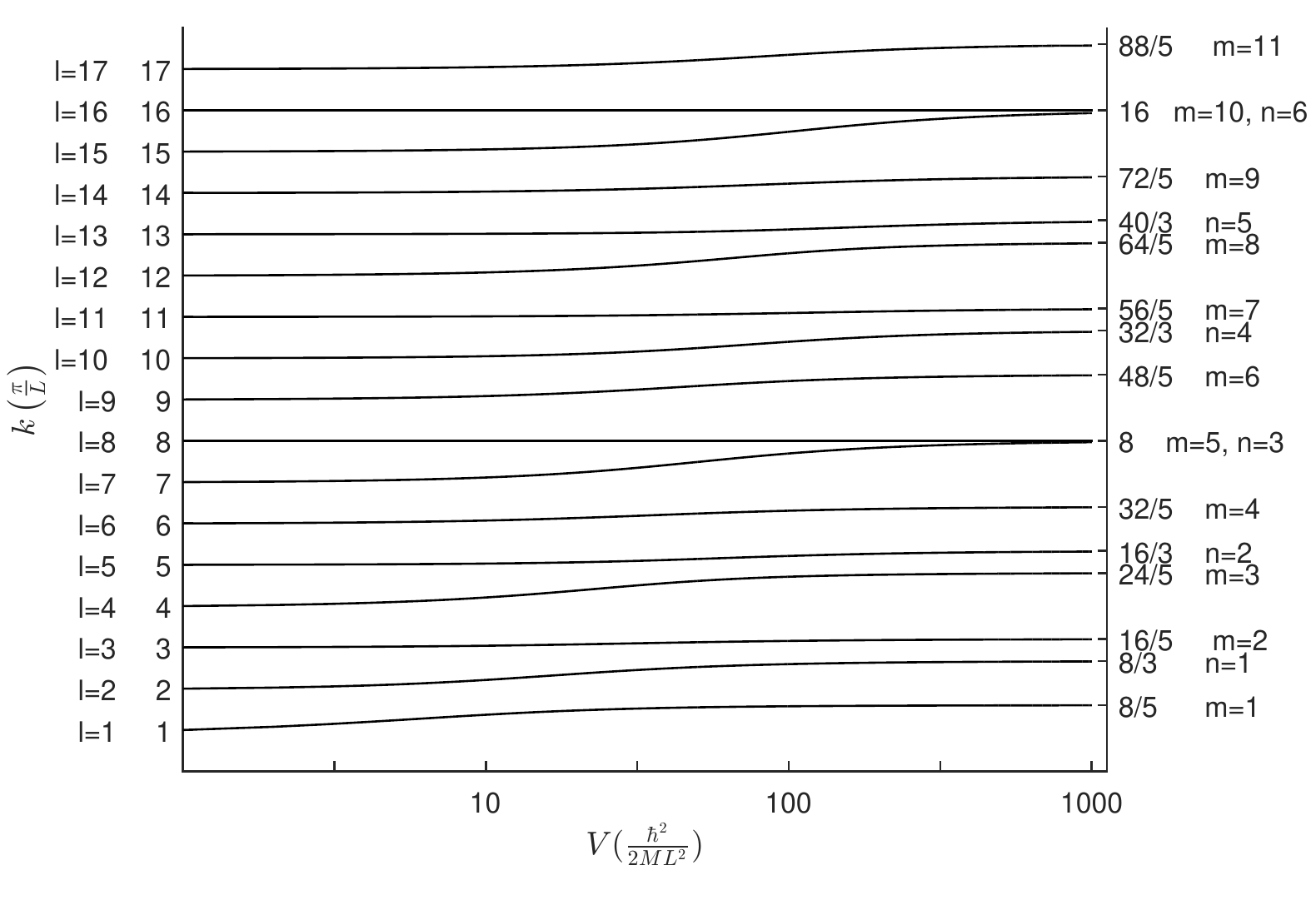}
\caption{Spectrum of the infinite square well with a delta-function potential of magnitude $V$ located at $x_0 = \frac{3}{8}L$. The levels are shown in terms of the wave number $k = \sqrt{2ME}$ and a logarithmic scale for $V$ (which has units of $[\hbar^2/2ML^2]$).}\label{KCurves} 
\end{center}
\end{figure}

\begin{figure}[H]
\begin{center}
\begin{tabular}{c}
\includegraphics[width=6in]{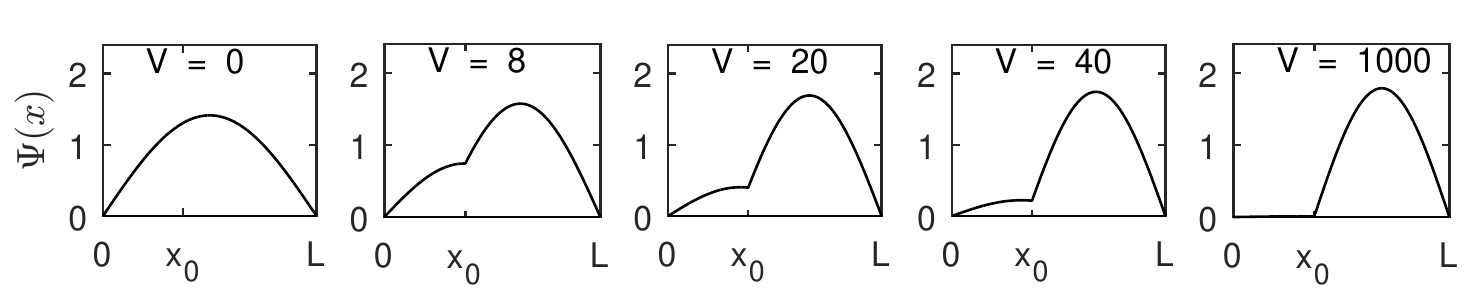} \\
\includegraphics[width=6in]{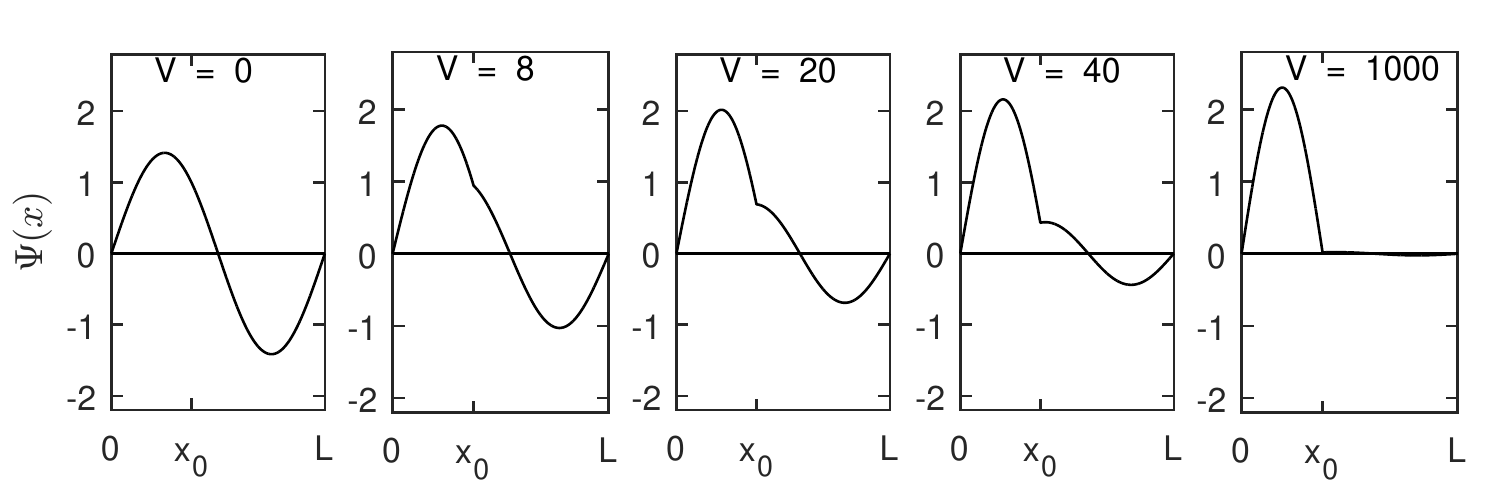} \\
\end{tabular}
\caption{The ground state (top) and first excited state (bottom) of the infinite square well with a delta-function potential of magnitude $V$ located at $x_0 = \frac{3}{8}L$. Clearly, as the barrier magnitude increases each eigenstate becomes confined on one side of the barrier or the other, becoming eigenstates of the individual wells.}\label{Ground_First}
\end{center}
\end{figure}

\begin{figure} [H]
\begin{center}
\begin{tabular}{c}
\includegraphics[width=6in]{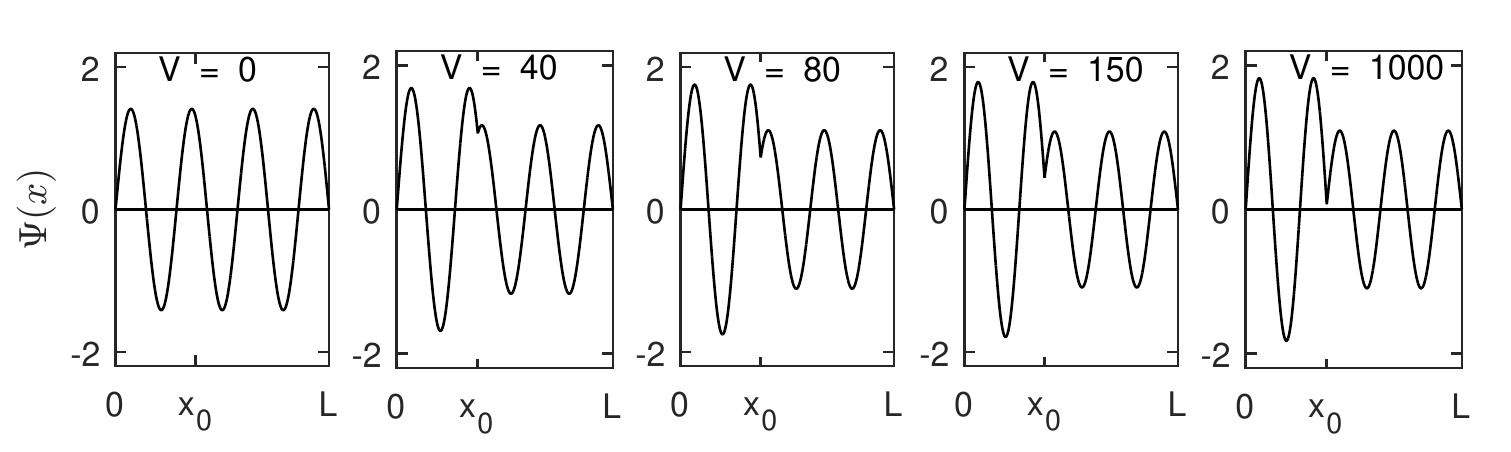} \\
\includegraphics[width=6in]{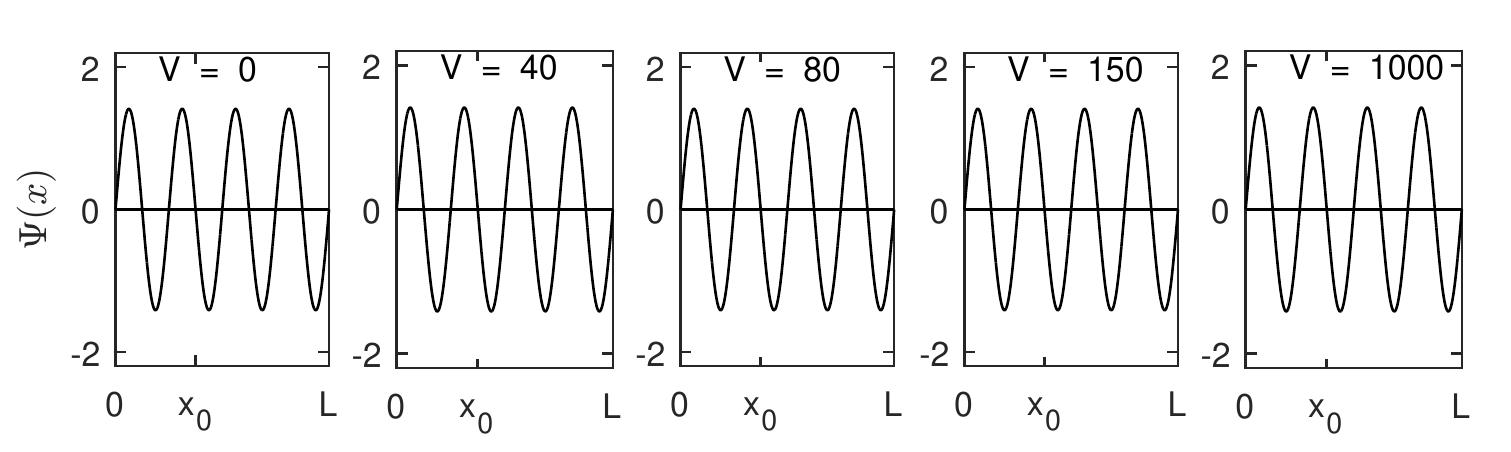} \\
\end{tabular}
\caption{The seventh (top) and eighth (bottom) excited state of the infinite square well with a delta-function potential of magnitude $V$ located at $x_0 = \frac{3}{8}L$. These eigenstates become degenerate and clearly fail to become confined on one side of the barrier or the other as the barrier magnitude increases, and thus they do not become eigenstates of the individual wells. Instead, the eigenstates of the individual wells are orthogonal superpositions of these two eigenstates of the original well, as shown in Equations (\ref{Left}) and (\ref{Right}).}\label{7th_8th}
\end{center}
\end{figure}

We have also considered the adiabatic limit in which a delta-function barrier at $x_0 = \frac{3}{8}L$ is raised very slowly, such that energy levels of the original un-split well transition gradually into the energy levels of the two wells after splitting. The shifting of the first seventeen energy levels is shown in Figure \ref{KCurves} in terms of the wave number $k = \sqrt{2ME}$. This figure and also Figures \ref{Ground_First} and \ref{7th_8th} were obtained by parametric solutions of the time-independent Schr\"{o}dinger equation, rather than a simulation of adiabatic evolution in time.

In general, as the barrier magnitude $V$ increases, many eigenstates of the well become gradually more and more confined to one side of the barrier or the other, with tunneling rates vanishing as the barrier becomes infinite. Figure \ref{Ground_First} shows the ground state and first excited state of the well as a function of the potential of the delta-function \mbox{barrie}.

However, this is not true of all eigenstates, which leads to some interesting physics for the cases where the two new wells have degenerate energy levels. In these cases, the adiabatic approximation fails, strictly speaking, because there will be a significant probability of transitions between the nearly-degenerate levels.

The $l=8$ energy level of the unsplit well is degenerate with the $n=3$ and $m=5$ levels of the two wells after splitting ($E_0(8) = E_1(3) = E_2(5)$), but a single energy level cannot divide into two orthogonal energy levels under adiabatic evolution. In fact, it is easy to see analytically that because the $l=8$ mode has a node at $x_0$, it will remain unchanged no matter how quickly or slowly the barrier is raised, meaning that it does not become confined to one side of the barrier, but rather becomes a superposition of the particle being on either side. As the barrier goes up, the $l=7$ mode gradually becomes degenerate with the $l=8$ mode, but also fails to become confined to one side of the barrier (see Figure \ref{7th_8th}). The $l=7$ mode develops a slope-discontinuity at $x_0$, such that the $l=8$ and $l=7$ eigenstates remain orthogonal, even as their energies become degenerate,
\[ \psi^0_8(x) = \sqrt{\frac{2}{L}}\left\{
\begin{array}{ll}
   \sin\frac{8 \pi x}{L} & 0 \leq x \leq x_0 \\
   -\sin\frac{8 \pi (L-x)}{L} & x_0 < x \leq L \\
\end{array}
\right. \]
\[ \lim_{V\rightarrow\infty} \psi^0_7(x) = A\left\{
\begin{array}{ll}
   \sin\frac{8 \pi x}{L} & 0 \leq x \leq x_0 \\
   \frac{3}{5}\sin\frac{8 \pi (L-x)}{L} & x_0 < x \leq L \\
\end{array}
\right. \]
where $A$ is a normalization constant.

Remarkably, since neither state becomes confined, it is not the case that the $l=7$ and $l=8$ eigenstates of the original well gradually become the $n=3$ and $m=5$ eigenstates of the two wells after splitting. Instead, the confined eigenstates of the two wells are superpositions of the these two degenerate states of the unsplit well. In the limit of infinite $V$, the left well $n=3$ eigenstate is,
\begin{equation}
\psi^1_3(x) = B \left( \psi^0_7(x) + \frac{3A}{5} \sqrt{\frac{L}{2}} \psi^0_8(x) \right), \label{Left}
\end{equation}
and the right well $m=5$ eigenstate is,
\begin{equation}
\psi^2_5(x) = C \left( \psi^0_7(x) - A \sqrt{\frac{L}{2}} \psi^0_8(x) \right), \label{Right}
\end{equation}
where $B$ and $C$ are new normalization constants.

For $x_0 = \frac{3}{8}L$, the same thing happens for all pairs of levels $l=8s$ and $l=8s-1$ for all integers $s\geq1$ and with the same coefficients in the superposition. In general, this effect happens whenever the specified degeneracy condition is present for any location of $x_0$ where a barrier is raised.

\subsection{Proposed Experiment}

The analysis of this paper can be applied equally well to a photon in a cavity, and this leads to a proposition for a simple experimental test of the ideas we present here, which would ultimately take the form of a type of frequency converter, similar to other work using superoscillations \cite{remez2015super}.

The idea is to use a square multimode fiber that acts as an infinite square well in two dimensions while being effectively free in the third dimension. If the superposition state of Equation (\ref{Psi}) can be created in one or both of the constrained dimensions, then there would be particular positions along the free dimension, corresponding to specific propagation times, for which the zero would be present in the wavefunction. A split in the fiber could begin at the location of such a zero, with the split now playing the role of a barrier that is raised very quickly inside the infinite square well, with the effective quickness coming from the propagation speed of the photon along the longitudinal direction.

\begin{figure}[H]
\begin{center}
\includegraphics[width=5.42in]{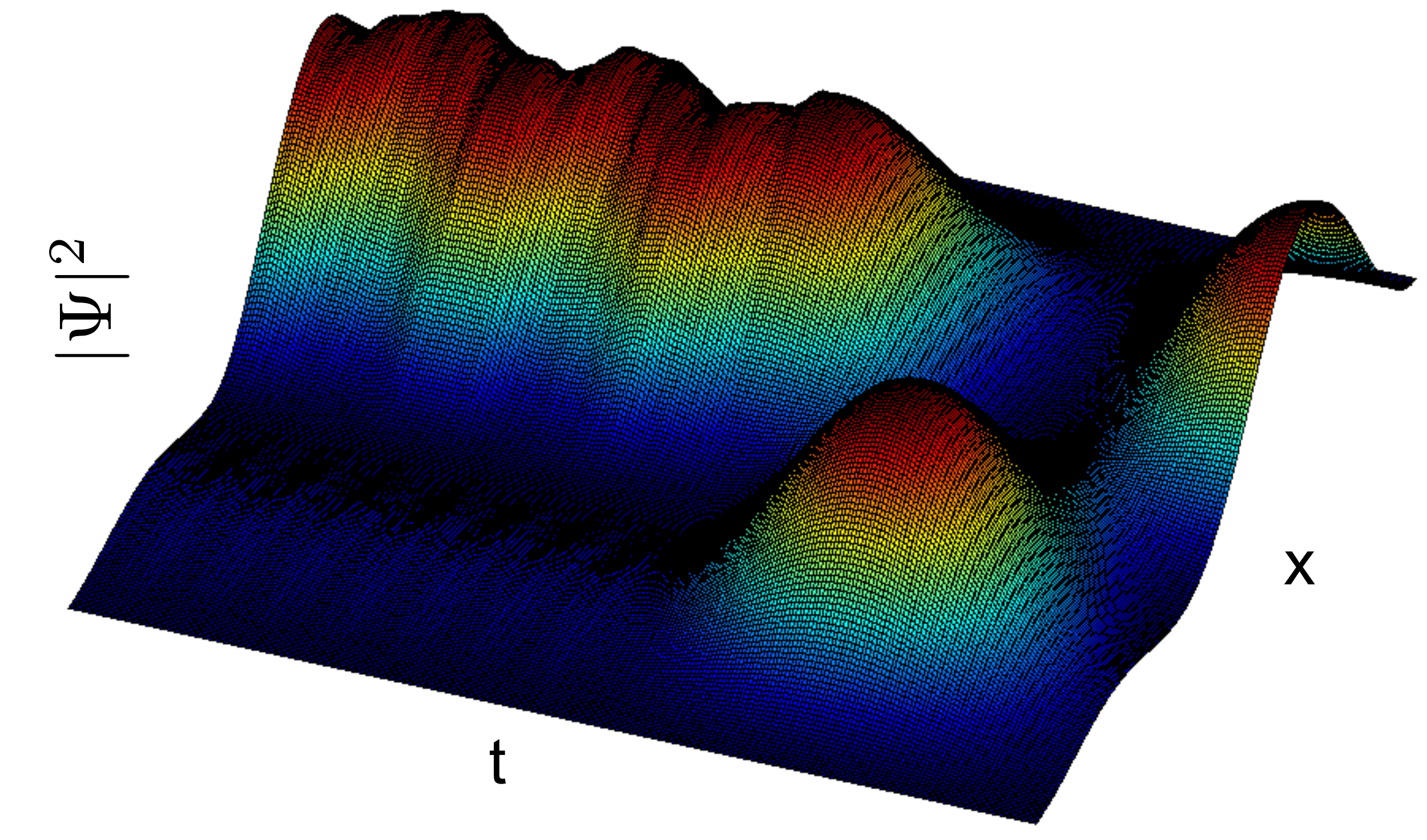}
\caption{The initial probability amplitudes for $|\psi(x,0)|^2$ appear on the right of the figure, and time evolves to the left. After one revival period, when the zero has reappeared at $x_0 = \frac{3}{8}L$, an infinite barrier is raised there, and the state evolves for the same period in the new potential.}\label{SpecialSplit} 
\end{center}
\end{figure}

Figure \ref{SpecialSplit} shows the time evolution $|\psi(x,t)|^2$ of the initial state $\psi(x)$ of Equation (\ref{Psi}), such that the state evolves within the original well for one revival period; then, the infinite barrier appears suddenly at $x_0 = \frac{3}{8}L$, and the state evolves for the same period in the new potential. The presence of higher energy modes with small amplitudes is clearly visible in the erratic behavior of $|\psi(x,t)|^2$ after the barrier is raised.

If our analysis is correct, then this should result in frequency conversion in the transverse mode(s) of the photon, such that the spectral superposition of a photon that emerges after the split will be changed, although the average energy is nearly unchanged. Components of the new superposition that remain inside the operating band of the multimode fiber will be incident on transverse frequency-sensitive detectors along the line, and components that are outside the band may escape, but could also potentially be captured by external detectors.

We should again stress that this frequency conversion is not in the direction of propagation along the fiber, but rather, it is the modes oscillating perpendicular to this direction that are converted.

Numerous technical details will need to be addressed before this experiment can be realized, but this goes beyond the scope of this paper.

\subsection{The Interference Energy Spectrum}

In this section, we develop the general theory of interference spectra for arbitrary superposition states. True superoscillatory functions are a special case of this phenomenon that contain stable zeros, which allow barriers to be raised much more slowly; although it remains unclear if they can be raised slowly enough to justify the adiabatic approximation.

Let us begin with a general superposition state of the 1-D
 square well,
\begin{equation}
\Psi(x,t) = \sum_{l=1}^\infty c_l \psi^0_l(x) e^{\frac{-i E_l t}{\hbar}}.
\end{equation}

This state has a unique set of zero points $S = \{(x_i, t_i)\}$, such that $\Psi(x_i,t_i) = 0$ and $0<x_i<L$ for all $(x_i,t_i) \in S$. At a given time $\tau$, it is possible to place infinite barriers at any or all points $(x_i,t_i)$ for which $t_i=\tau$ in order to split the well into smaller wells.

Let us suppose that at time $\tau$, we choose to divide the well into $N+1$ smaller wells by placing barriers at a set of $N$ available zeros from $S$, $\{\chi_j\}$ with $j=1,...,N$. We append the two endpoints $\chi_0=0$ and $\chi_{N+1}=L$ to this set and expand the index, such that $\chi_j>\chi_{j-1}$ is defined for all $j=1,...,N+1$. The resulting interference energy spectrum that is now available is,
\begin{equation}
E_j(k_j) = \frac{\hbar^2 \pi^2 k_j^2}{2M(\chi_j-\chi_{j-1})^2},
\end{equation}
for wells $j=1,...,N+1$, where $k_j = 1,2,...,\infty$ is the quantum number of the $j$-th well

In general, the union set of all possible spectra that can be obtained by placing any number of barriers at any number of points $(x_i,t_i)$ in $S$ is the {\it complete interference spectrum} of the state $\Psi(x,t)$.

We now switch to the Hilbert space representation of this problem in order to give the calculated probabilities, noting that the domain of each well, $(\chi_{j-1},\chi_j)$, is a different Hilbert space. We introduce the ket $|0\rangle$ to represent the original well and a $d$-level quantum system with basis $\{|j\rangle\}$ to represent the different split wells, with $d=N+1$. Each well is an infinite-dimensional Hilbert space, but the projectors onto each well have {\it relative rank} (subscript) proportional to the well size, such that,
\begin{equation}
|0\rangle\langle 0|_L \equiv \sum_{j=1}^{N+1} |j\rangle\langle j|_{\chi_j-\chi_{j-1}}.
\end{equation}

To begin, we represent $\Psi(x,\tau)$ as,
\begin{equation}
|\Psi\rangle = \sum_{l=1}^\infty d_l |l\rangle |0\rangle = \sum_{l=1}^\infty d_l \sum_{j=1}^{N+1} |l_j\rangle |j\rangle,
\end{equation}
where $|l\rangle$ represents ${\psi^0_l(x)}$, and $|l_j\rangle$ is the unit ket for the (unnormalized) truncated eigenfunction in the $j$-th well, such that $\{|j\rangle|l_j\rangle\}$ is a complete orthonormal basis.

We compute the similarity matrices $\{\hat{A}_j\}$ that perform the change of basis on each well,
\begin{equation}
\hat{A}_j = |j\rangle\langle j| \sum_{k_j=1}^\infty \sum_{l=1}^\infty A_{k_j l} |k_j\rangle\langle l_j|, \label{MatrixEls}
\end{equation}
where $|k_j\rangle$ are the normalized energy eigenstates of the $j$-th well, corresponding to,
\begin{equation}
\psi^j_{k_j}(x) = \sqrt{\frac{2}{\chi_j - \chi_{j-1}}} \sin\left[ \frac{k_j \pi (x-\chi_{j-1})}{\chi_j - \chi_{j-1}} \right],
\end{equation}
and the matrix elements are given by,
\begin{equation}
A_{k_j l} = \langle k_j|l_j\rangle = \int_{\chi_{j-1}}^{\chi_j} \frac{2}{\sqrt{L(\chi_j - \chi_{j-1})}}   \sin\left[ \frac{k_j \pi (x-\chi_{j-1})}{\chi_j - \chi_{j-1}} \right] \sin\left[\frac{l \pi x}{L}\right] dx
\end{equation}
\begin{equation}
=\frac{2 k_j L}{\pi}\sqrt{L(\chi_j - \chi_{j-1})}\left[\displaystyle\frac{(-1)^{k_j}\sin\left(\displaystyle\frac{l \pi \chi_j}{L}\right) - \sin\left(\displaystyle\frac{l \pi \chi_{j-1}}{L}\right)}{l^2(\chi_j-\chi_{j-1})^2-k_j^2L^2}\right].
\end{equation}

Finally, letting the matrices $\hat{A_j}$ act on the wavefunction, we obtain the representation in terms of the new energy eigenbasis:
\begin{equation}
\sum_{j=1}^{N+1} \hat{A}_j |\Psi\rangle = \sum_{j=1}^{N+1} |j\rangle\langle j| \left[\sum_{k_j=1}^\infty \sum_{l=1}^\infty A_{k_j l} |k_j\rangle\langle l_j| \right] \sum_{l'=1}^\infty d_{l'} \sum_{j'=1}^{N+1} |l'_{j'}\rangle |j'\rangle
\end{equation}
\begin{equation}
= \sum_{j=1}^{N+1} |j\rangle \sum_{k_j=1}^\infty \sum_{l=1}^\infty d_l A_{k_j l} |k_j\rangle.
\end{equation}

In this form, it is clear that this is an entangled state in the sense that the energy levels are correlated to specific wells. If we project onto a particular well $|j\rangle$, we get a superposition of the energy eigenstates of that well. Likewise, if we project onto a specific energy eigenstate $|k_j\rangle$, we are also projecting onto a specific well (or several wells with degenerate energies).

The probability of finding the particle in the $j$-th well with energy $E_{k_j}$ is,
\begin{equation}
P(k_j) = \left|\sum_{l=1}^\infty d_l A_{k_j l} \right| ^2,
\end{equation}
and the total probability of finding it in each well is,
\begin{equation}
P(j) = \int_{\chi_{j-1}}^{\chi_j} |\Psi(x)|^2 dx = \sum_{k_j=1}^\infty \left|\sum_{l=1}^\infty d_l A_{k_j l} \right| ^2.
\end{equation}

\section{Discussion}

The results given here raise interesting questions about the conservation of energy. Unlike the genuine superoscillating functions that have been studied elsewhere, the method we are using to split the wells requires that the barrier be raised very quickly, and because the $\Delta t$ of this process is so small, it necessarily introduces a large $\Delta E$ due to the Heisenberg uncertainty principle. Thus, even though $\langle E \rangle$ is unchanged by the sudden addition of the barrier at a zero of the wavefunction, the change in energy between the individual pre-barrier and post-barrier levels can be supplied by the $\Delta E$. Therefore, in this case, the uncertainty principle washes out any possibility that the conservation of energy is violated.

In general, one may consider the condition $\Delta \langle E \rangle = 0$ sufficient to show that energy conservation is obeyed. Even though the Hamiltonians are different before and after the barrier is present, the complete eigenbases of the two Hamiltonians both span the space of normalizable functions on the interval $x\in[0,L]$ with zeros at $x=\{0, x_0, L\}$. If $|\psi\rangle$ is a true superposition state, then there is nothing unexpected about finding energies $E_1(n)$ or $E_2(m)$ when the energy is measured in the eigenbasis $\{ |n\rangle, |m\rangle \}$. The potential problem arises if one assumes that the discrete spectrum $E_0(l)$ is an inherent property of the state $|\Psi\rangle$, and it should be impossible to measure other energies.

 Supposing there is an inherent preferred spectrum, then as the barrier goes up, the energy levels of the original well divide and smoothly transition to the energy levels of the split well. The particle only interacts with the barrier, and so, by energy conservation, it must be the case that the change in energy is supplied by the barrier. If the particle is found with energy eigenvalue $k_j$ in the split well, it has probability $P_{k_j}(l)$ to have transitioned from energy eigenvalue $l$ of the original well, for which the barrier must have supplied the energy change $\Delta E_{k_jl} = E_j(k_j) - E_0(l)$. We thus define the barrier state $|B_{k_j l}\rangle$ as the state in which the barrier lost this energy. Each split-well energy eigenstate of the joint particle-barrier system $PB$ is then of the form,
\begin{equation}
|k_j\rangle\langle k_j|^{PB} = \sum_{l=1}^\infty P_{k_j}(l) |l_j\rangle\langle l_j||B_{k_j l}\rangle\langle B_{k_j l}|.
\end{equation}

Thus, we see that after the barrier is raised, the state $|\Psi\rangle$ is entangled with the barrier through energy conservation. The amount of free energy needed to produce the individual shifts $\Delta E_{k_jl}$ is easily provided by the large uncertainty $\Delta E$ for a sudden barrier.

It should be possible to experimentally test this interaction by performing an ensemble of runs of the experiment and taking the average of all barrier-energy measurements conditioned on post-selecting a particular energy eigenstate $|k_j\rangle$ of the particle in the split well. This average should wash out the noise introduced by the large $\Delta E$, making it possible to measure the average \mbox{barrier energy},
\begin{equation}
\langle E^B \rangle = -\sum_{l=1}^\infty P_{k_j}(l)\Delta E_{k_jl}.
\end{equation}

 Now, there are several viable interpretations of how $P_{k_j}(l)$ should be defined if we wish to assume a preferred spectrum $|l\rangle$ as an inherent property of the superposition state: a property not specified by the wavefunction alone.

 One simple choice is $P_{k_j}(l) = |d_l|^2$, which means that regardless of which $|k_j\rangle$ the state is found in, the probability that it transitioned from energy level $E_0(l)$ is the same as the probability to find the state with that energy in the original well. Using this form for our simple example case with $x_0 = 3L/8$ and the initial superposition state of Equation (\ref{Psi}), the probability to find the particle in the ground state of the smaller well, with energy $(8/3)^2 E_0(1)$, is roughly 6\%, and the average energy of the barrier, post-selected on this outcome, is $\langle E^B \rangle = -4.22 E_0(1)$. This case is particularly interesting because the wavefunction is zero at the location of the barrier, and yet, the particle and barrier seem to exchange significant quantities of energy as the barrier is raised.

 Alternatively, we can consider a quasi-probability treatment of the superposition state. Given the initial state $|\Psi\rangle$ corresponding to Equation (\ref{Psi}) and the final state $|k_j\rangle$, the best estimate for the probability that the intermediate state was $|l\rangle$ is given by the weak value of the projector $|l\rangle\langle l|$ \cite{dressel2015weak, hall2001exact, johansen2004value, hall2004prior},
\begin{equation}
\tilde{P}_{k_j}(l) = \Re \frac{\langle k_j |l\rangle\langle l|\Psi \rangle}{\langle k_j |\Psi \rangle} = \Re \frac{A_{k_j l}d_l}{\sum_{l'=1}^\infty A_{k_j l'}d_{l'}}.
\end{equation}

We should stress that this quasi-probability can be less than zero or greater than one, which makes its physical interpretation somewhat unclear. Indeed, in our example case, post-selecting on the ground state of the smaller well, we find that the quasi-probabilities that the previous energy level was $l=1$ or $l=2$ are $\tilde{P}_{k_1}(1) = -1.037$ and $\tilde{P}_{k_1}(2) = 2.037$, respectively. Nevertheless, if we use $|k_j\rangle\langle k_j| = \sum_{l=1}^\infty \tilde{P}_{k_j}(l) |l_j\rangle\langle l_j||B_{k_j l}\rangle\langle B_{k_j l}|$ after the barrier is raised, then the average energy of the barrier, post-selected on this outcome, is $\langle E^B \rangle = 0$, which is quite a striking result, that turns out to be completely general, as we now show.

For a general superposition state $\Psi(x) = \sum_{l=1}^\infty d_l \sqrt{\frac{2}{L}} \sin \frac{l \pi x}{L}$, the condition that $\Psi(x)$ has a zero at $x_0$ is simply $\sum_{l=1}^\infty d_l \sin \frac{l \pi x_0}{L} = 0$. If a barrier is placed at $x_0$ for a general wavefunction and the particle subsequently collapses into eigenstate $|k_1\rangle$ of the well between $x=0$ and $x=x_0$, the average energy of the barrier using the quasi-probability, $\tilde{P}_{k_1}(l)$, is,
\begin{equation}
\langle E^B_{k_1} \rangle = \frac{\sum_{l=1}^\infty d_l \sin \frac{l \pi x_0}{L}}{\sum_{l'=1}^\infty \frac{d_{l'}}{\Delta E_{k_1l'}} \sin \frac{l' \pi x_0}{L}}.
\end{equation}

Thus, we see that for any initial wavefunction with a zero at $x_0$, the average energy of a barrier that is suddenly raised at $x_0$, conditioned on the post-selection of any one outcome $|k_1\rangle$ in the split well, is always zero. We used a quasi-probability distribution to obtain this result, but we are not actually predicting that any physical event occurs with probability $\tilde{P}_{k_1}(l)$; rather, it is used as an intermediate calculation tool to address the fact that the initial state was a superposition of multiple $|l\rangle$ eigenstates. Quasi-probabilities outside the range zero to one are also known to be related to quantum contextuality \cite{pusey2014anomalous,spekkens2008negativity}.

There is one other definition one might use for $P_{k_j}(l)$, which are the probabilities for the mixed state prepared as $\rho = \sum_{l=1}^\infty |d_l|^2 |l\rangle \langle l |$. This is the least physical choice, because the quantum interference terms have been removed, and the predicted probabilities of different outcomes are entirely different than for $|\Psi\rangle$. Nevertheless, $\rho$ does explicitly have the energy spectrum $|l\rangle\langle l|$, which is unclear for $|\Psi\rangle$, and thus, it may be relevant here. We can obtain,
\begin{equation}
P_{k_j}(l) = \frac{|A_{k_j l}|^2|d_l|^2}{\sum_{l'=1}^\infty |A_{k_j l'}|^2|d_{l'}|^2},
\end{equation}
and using these probabilities in our example case, we obtain $\langle E^B \rangle = -3.73 E_0(1)$. In this case, the exchange of energy between the particle and barrier can be explained by local interaction, since the individual $|l\rangle$ are not generally zero at $x_0$.

Regardless of which explanation we use for the shifts of individual levels, the wavefunction is unchanged, and the overall average energy provided by the barrier is still zero. This raises interesting questions about the physics of measuring any quantum state in an alternate eigenbases, even a spin. If the preferred spectrum (both eigenvalues and eigenstates) is an inherent property of the quantum state of a system, then changing the basis prior to a measurement has a direct effect on that inherent property, but without changing the state vector itself (\emph{i.e.}, without collapse). This would then be an explicit manifestation of quantum contextuality \cite{KS, spekkens2005contextuality}, in that the choice of measurement context physically changes an internal property of the state: the preferred basis.

Conducting the experiment to measure the barrier energy may prove technically challenging, but it would allow us to test our suppositions for the form of ${P}_{k_j}(l)$.

On the other hand, taking the viewpoint that the state has no inherent discrete energy spectrum (and thus, ${P}_{k_j}(l)$ is meaningless) and noting that the barrier cannot interact locally with the particle, which has zero probability to be found in the same place as the barrier, then the average energy of the barrier should be $\langle E^B \rangle=0$, regardless of post-selection.

It is interesting that even this experiment cannot distinguish the case of an inherent preferred spectrum with a quasi-probability distribution from the case of no inherent spectrum at all, since both predict $\langle E^B \rangle=0$.

Finally, if we consider the case of a superoscillating wavefunction \cite{aharonov2016super}, the barrier can be raised very slowly at quasi-stable nodes of the superoscillation, and the large $\Delta t$ leads to a small $\Delta E$, small enough that it may not be enough to encompass the individual shifts $\Delta E_{k_jl}$. Thus, if we insist that the wavefunction has an inherent preferred spectrum, we may be presented with a violation of conservation of energy, especially if $\langle E^B \rangle\neq0$. If $\langle E^B \rangle=0$, then it may still be that the barrier simply facilitates exchanges of energy between different levels of the particle, in which case its own $\Delta E$ may not matter. On the other hand, if the wavefunction has no inherent preferred spectrum, there are no shifts $\Delta E_{k_jl}$, and the issue vanishes.


\section{Materials and Methods}

The details of our numerical simulation of the time-dependent Schr\"{o}dinger equation can be found in the Appendix. The simulation was written in MATLAB 2014a, and the code and data are available upon request.


\section{Conclusions}

We have explored the idea, and verified through simulation, that the energy eigenbasis of a state of the infinite square well can be altered through the addition of sudden potential barriers without a change to the state or its average energy. We consider the interpretation that this is a measurement in an alternate energy eigenbasis, because the energy eigenstates onto which the particle can collapse after the barrier is raised are different than the eigenstates of the original well and have different energies. The main point of contention with this view is that it is common to interpret a superposition state of the infinite square well as having a discrete list of preferred spectral energies as an inherent property, such that it is impossible to measure any energy not on this list. This is inconsistent with the idea that a genuine superposition state can be measured in different bases and can collapse onto any eigenstate in the measured basis. If we ascribe an inherent discrete spectrum directly to the superposition state, then the barrier must exchange energy with the particle in order to adjust that spectrum. This interaction must occur despite the fact that the particle has zero probability to be found at the location where the barrier is raised ($|\psi(x_0)|^2=0$), but this may be consistent, because the discrete energy spectrum of a bound particle is not usually considered a local property of the wavefunction; thus, the inherent spectrum must be a de-localized, or possibly nonlocal, property of the quantum state of the particle. In principle, an experiment might allow us to determine if this inherent preferred spectrum exists by measuring the barrier energies post-selected on obtaining particular measurement outcomes.

The alternative viewpoint, that the state has no inherent discrete spectrum, raises interesting questions about dynamical collapse, since it implies that the particle must somehow probe the entire shape of its binding potential as part of the dynamical collapse process, in order to cause a collapse into an eigenstate of the correct Hamiltonian.

Following this research, we have continued to explore the idea that there is no preferred discrete energy spectrum inherent to a wavefunction at all; but rather, it is always the measurement Hamiltonian that determines the spectrum, and this is where quantization appears. The wavefunction itself is not quantized, and its evolution can be modeled by considering its Fourier transform into a continuous spectrum of plane waves. While this work is ongoing, one preliminary result of some interest is the fact that for certain states, there are discrete zeros in the Fourier transform of the state, which means that in any discrete energy eigenbasis that can be used to measure the state, the probability of obtaining that spectral energy is zero. Thus, while the allowed energy levels of such wavefunctions are not generally quantized, the forbidden energies {\it are} quantized. We call this amusing phenomenon {\it unquantum mechanics}. The only states that do have discrete quantum spectra in the Fourier transform domain are unnormalizable continuous plane waves. We plan to develop these ideas further in a subsequent paper.

We have found little in the literature that seems specifically relevant to the new ideas presented here, so we provide a collection of citations on work that is somewhat more distantly related in order to flesh out the state of the art. These topics include, frequency conversion using nonlinear optics and other systems, double-well potentials in Bose--Einstein condensates and other systems \cite{zin2006method, mahmud2002bose, shin2004atom, schumm2005matter, jaaskelainen2005dynamics,infeld2006statics, spekkens1999spatial, bavli1992laser, kierig2008single}, energy-time uncertainty, and superoscillations \cite{aharonov2012superoscillation2,aharonov2011some,aharonov2013cauchy,aharonov2015superoscillating1,aharonov2015mathematics, aharonov2015superoscillating2, buniy2014quantum}.\newline\newline

\textbf{Acknowledgments:} We would like to thank Sandu Popescu, Michael Berry, Justin Dressel, Matthew Leifer and \mbox{Jeff Tollaksen} for helpful conversations as this research took shape. This research was supported (in part) by the Fetzer-Franklin Fund of the John E. Fetzer Memorial Trust. This work has been supported in part by the Israel Science Foundation Grant No. 1311/14.  The authors would also like to acknowledge the use of the Samueli Laboratory in Computational Sciences in the Schmid College of Science and Technology, Chapman University, for the computers we used for the simulation.

\appendix

\section*{\noindent Appendix: Numerical Simulation} \label{Simulation}

In order to characterize the effect of rapidly raising a potential barrier within the infinite square well, we numerically solve the time-dependent Schr\"odinger equation during the period $\tau$ that the potential is changing \cite{becerril2008solving, zin2006method}. We used a modified version of the fourth order Runge--Kutta method, where the modification is a first order approximation in the size of the time steps $\Delta t$ used in the simulation. This approximation leads to an update at each time step of,
\begin{equation}
\psi(x,t+\Delta t) = \psi(x,t) + i\Delta t\left(\frac{\partial^2}{\partial x^2} - \frac{1}{6}[V(x,t) + 4V(x,t+\Delta t/2) + V(x,t+\Delta t)]\right) \psi(x,t),
\end{equation}
where we are working in units with $\hbar=1$ and $2M=1$. The first order approximation is validated by the fact that for the range of parameters values for which we run the simulation, $\left|\Delta t\left(\frac{\partial^2}{\partial x^2}-\frac{1}{6}[V(x,t) + 4V(x,t+\Delta t/2) + V(x,t+\Delta t)] \right)\psi(x,t)\right| \ll |\psi(x,t)|$.

The magnitude of $\psi$ is so large compared to the correction that significant numerical round-off error can be introduced. To avoid this, the correction terms of different magnitudes are accumulated as separate variables in the simulation and added together freshly in order to compute the correction~at the each time step. This also allows precise computation of the average energy change, as shown~below.

For the time-dependent barrier, we use a linear rate of increase and a normalized dimensionless Gaussian kernel of full width $w$ at half-maximum,
\begin{equation}
G_w(x) = \frac{e^{-4 (\ln{2}) \left(\frac{x-x_0}{w}\right)^2}} { \int_0^L e^{-4 (\ln{2}) \left(\frac{x-x_0}{w}\right)^2}dx},
\end{equation}
which reduces to the usual general form,
\begin{equation}
G_w(x) \approx \frac{2}{w} \sqrt{\frac{\ln{2}}{\pi}}e^{-4 (\ln{2}) \left(\frac{x-x_0}{w}\right)^2},
\end{equation}
for narrow widths. This last form also becomes the Dirac delta-function in the limit that $w \rightarrow 0$.
We define the parameter $A_w$ as the overlap between the probability distribution of the initial state $\psi(x,0)$ and the normalized barrier kernel,
\begin{equation}
A_w = \int_0^L |\psi(x,0)|^2 G_w(x) dx,
\end{equation}
which we will use later.

The potential is then,
\begin{equation}
V_w(x,t) = \frac{t}{\tau}V_m G_w(x),
\end{equation}
for $0 \leq t \leq \tau$, and $V_m=10^4 \frac{\hbar^2}{2ML^2}$. As discussed above, this $V_m$ is more than sufficient to cause the desired splitting of the lowest energy levels and eigenstates.

We set $x_0 = \frac{3}{8}L$ and $L=1$, and divided the simulation into $10^3$ time steps, which gives $\Delta t = 10^{-3}\tau$. We performed the simulation using the method of lines, with a mesh of $x$ values from zero to one with step size $\Delta x = 10^{-5} L$. The simulation preserves the normalization of $\psi(x,t)$ to very high precision, with the largest normalization error on the order of $10^{-9}$. This serves as an estimate of the numerical tolerance of the simulation and provides some verification that it is \mbox{working correctly}.

At the end of the model, the original state $\psi_0(x) \equiv \psi(x,0)$ has evolved into the state \mbox{$\psi(x,\tau) = \psi(x,0) + \Delta\psi(x)$}. At time $t=0$, the expectation value of the potential energy is \mbox{$\langle V_0\rangle = \int_0^L |\psi(x,0)|^2 V(x,0) dx = 0$}, and kinetic energy is $\langle K_0\rangle = \frac{\hbar^2}{2m}\int_0^L \psi^*(x,0) \psi''(x,0) dx$.

At time $t=\tau$, the expectation value of the potential energy is,
\begin{equation}
\langle V_\tau \rangle = \int_0^L \left( |\psi_0(x)|^2 + |\Delta\psi(x)|^2 + \psi^*_0(x)\Delta\psi(x) + \Delta\psi^*(x)\psi_0(x) \right)V(x,\tau)dx,\label{V}
\end{equation}
which is also equal to the change in the potential energy. We separate this into different terms: $\Delta\langle V \rangle_{\psi_0} = V_m\int_0^L |\psi(x,0)|^2 G_w(x) dx = A_w V_m$ is the change in potential energy simply due to the ``lifting'' of the initial state. Note that $\Delta\langle V \rangle_{\psi_0}$ clearly approaches zero for a very narrow barrier centered at a zero of the $\psi(x,0)$. The last three terms in Equation (\ref{V}) are due to the change in the state, and we call the sum of these terms $\Delta \langle V\rangle_{\Delta\psi} $.

At time $t=\tau$, the expectation value of the kinetic energy is,
\begin{equation}
\langle K_\tau \rangle = -\frac{\hbar^2}{2m} \int_0^L \psi^*_0(x)\psi''_0(x) dx
\end{equation}
\begin{equation}
- \frac{\hbar^2}{2m} \int_0^L \left(\Delta\psi^*(x)\Delta\psi''(x)+ \psi^*_0(x)\Delta\psi''(x) + \Delta\psi^*(x)\psi''_0(x) \right) dx.\label{K}
\end{equation}

In this case, the first term is simply the initial kinetic energy $\langle K_0 \rangle$, and the other three terms are the change in the average kinetic energy $\Delta\langle K \rangle$.

We ran the simulation for a wide range of dimensionless parameters (shown here with corresponding physical units). The domain of the search was all combinations of the following choices of initial parameters:
\begin{itemize}
\item $\psi_0 = \sqrt{\frac{2}{L}} \sin \frac{\pi x}{L}, \sqrt{\frac{2}{L}} \sin \frac{2\pi x}{L}, \sqrt{\frac{2}{L}} \sin \frac{3\pi x}{L} $, $\sqrt{\frac{2}{L (\alpha^2+1)}}\left(\alpha \sin\frac{\pi x}{L} \pm \sin\frac{2 \pi x}{L}\right)$ (with $\alpha \equiv 2\cos\frac{\pi x_0}{L} = \sqrt{2-\sqrt{2}})$
\item $\tau = \{ 10^{-10}, 10^{-11}, 10^{-12}, 10^{-13}, 10^{-14}\} [{2ML^2}/{\hbar}]$
\item $w = \{ (1,2,3,4,5,6,7,8,9)\times 10^{-4}, (1,2,3,4,5,6,7,8,9)\times 10^{-3},(1,2,3,4,5,6,7,8,9)\times 10^{-2}, (1,2,3,4,5,6,7,8,9)\times 10^{-1}, (1,2,3,4,5,6,7,8,9,10) \} [L]$
\end{itemize}

The two $\psi_0$'s denoted by the $\pm$ are the state with a zero at the center of the barrier and its reflection, which is maximum at the center of the barrier. The narrowest barrier width we can reasonably simulate with a mesh spacing of $10^{-5}$ is $w=10^{-4}$, while for $w=10$, the barrier is approximately flat, and the bottom of the entire well is raised uniformly.

We do not have analytic forms for $\Delta \langle V \rangle_{\Delta\psi}$ or $\Delta \langle K \rangle$ for all choices of the parameters $V_m$, $w$, $\tau$ and $\psi_0$ in the simulation, but we were able to deduce reasonably good fits for the narrow-barrier regime ($w \leq 10^{-2} [L]$) by guessing that these would be separable into a product of functions of each parameter individually. In this regime, the dependence on $\psi_0$ can be reduced to a simple dependence on the derived quantity $A_w$, but for larger widths, there is more direct dependence on $\psi_0$.

The forms we obtain for the separate functions are somewhat odd, but they are in terms of dimensionless quantities and provide reasonable fits to the simulated data. We found using logarithmic analysis that the function obeys power laws in its various parameters. We do not ascribe any analytic meaning to these forms.

The fit functions in terms of the dimensionless parameters are,
\begin{equation}
\Delta \langle K \rangle^\textrm{fit}(\tau, w, \psi_0, V) = C_K\frac{A_w \tau^2V^4}{ w^{3}}, \label{DKfit}
\end{equation}
with $C_K = 4.9895 \times 10^{-9}$ and
\begin{equation}
\Delta \langle V \rangle_{\Delta\psi}^\textrm{fit}(\tau, w, \psi_0, V) = C_V\frac{A_w \tau^2 V^p}{ w^q}, \label{DVfit}
\end{equation}
with $C_V = -1.5921\times10^{-7}$, $p = 4.3164$ and $q=2.3146$. These functions also show the time dependence through $V(t) = \frac{t}{\tau}V_m$.
Because they range over many order of magnitude, we compute the error of these functions by finding the root-mean-square error between the logarithms of the simulation data and the fit for all combinations of parameters in the narrow-barrier regime. This then gives us the relative errors, $\delta\Delta \langle K \rangle^\textrm{fit} = 16.21\%$ and $\delta\Delta \langle V \rangle_{\Delta\psi}^\textrm{fit} = 57.05\%$. Clearly, the fit is much better for the kinetic energy, but in both cases, we see that the fit functions provide good estimates of the orders of magnitude of $\Delta \langle K \rangle$ and $\Delta \langle V \rangle_{\Delta\psi}$, which is sufficient for our purpose here.

It is important to note that in order to use this fit function, one must put $w$ into units $[L]$, $\tau$ in units of $[{2ML^2}/{\hbar}]$ and $V$ in units of $[{\hbar^2}/{2ML^2}]$, and the output will be in units of $[{\hbar^2}/{2ML^2}]$. The numerical values are then dimensionless.

We interpret the positive curve in $\Delta \langle K \rangle$ as the barrier goes up as kinetic energy imparted to the particle by the barrier. This kinetic energy appears as the wavefunction is pushed away from the barrier on both sides, and this also results in the negative curve for $\Delta \langle V \rangle_{\Delta\psi}$.

Overall, the simulation shows that any change in the potential, performed sufficiently quickly, will have a negligible effect on the wavefunction or its energy. The change in energy is actually largest for the narrowest barriers and goes to zero in the wide limit, where the entire flat bottom of the well is being raised. We see that in order to minimize the change in energy, narrower barriers must be raised faster and located at zeros of $\psi(x)$ where $A_w$ will be minimized. It is important to note that for the state in Equation (\ref{Psi}), with a zero at $x_0$, we find that $A_w \propto w^2$, and so, even in this case, the energy change appears to go to infinity as $w \rightarrow 0$ for a finite $\tau$. In principle, one can split a state with a wide flat zero, for which $A_w=0$ up to some $w$, but such a state would contain significant contributions from high-energy modes and may fall outside the regime where our first order approximation in time steps is valid.

As discussed above, our simulation shows that it is quite possible to accomplish the desired splitting of the lowest energy levels of the well without significantly altering the kinetic energy of the state, by using Gaussian barriers with physically-plausible widths and speeds. In general, the change in the wavefunction is very small, and by far, the largest effect on the energy of the state is due to the ``lifting'' of the original state, $\Delta\langle V \rangle_{\psi_0} = A_w V_m$.

For the example discussed above, with $w=10^{-3}[L]$, $V_m=10^4 [\hbar^2/2ML^2]$, $\tau = 10^{-10}[2ML^2/\hbar]$ and $\langle K_0 \rangle = 2.8918 \pi^2 [\hbar^2/2ML^2]$, the simulation returns,
\begin{equation}
\Delta \langle K \rangle / \langle K_0 \rangle \approx 1.58 \times 10^{-10} ,
\end{equation}
\begin{equation}
\Delta \langle V \rangle_{\Delta\psi} /\langle K_0 \rangle \approx - 5.19 \times 10^{-10} ,
\end{equation}
and:
\begin{equation}
\Delta\langle V \rangle_{\psi_0}/\langle K_0 \rangle \approx 2.29 \times 10^{-3} .
\end{equation}

Thus, it is clear that $\Delta \psi$, and thus, $\Delta \langle K \rangle$ and $\Delta \langle V \rangle_{\Delta\psi}$ are negligible. There is a noticeable (but small) increase in potential energy, $\Delta\langle V \rangle_{\psi_0}$, due to the lifting of the initial wavefunction. This can be treated as a correction when considering conservation of energy; the kinetic energy of the wavefunction is our \mbox{primary interest}.

We conclude by noting that the simulation we developed can be applied with many other choices of parameters. In particular, we could use many different barrier shapes other than Gaussian, and the barrier need not be raised linearly in time. We could also consider a much broader set of initial wavefunctions. Our goal here was to show the behavior as the barrier approaches a delta function at a zero of the wavefunction, and we believe that the parameter set we used and the fits we obtained for the narrow-barrier regime were quite sufficient to that task.

\bibliographystyle{ieeetr}
\bibliography{Interference_Spectrum_arXiv.bbl}

\end{document}